\documentclass[twoside]{article}
\usepackage{qic,epsfig}
\usepackage{hyperref}
\usepackage{tikz}					
\usetikzlibrary{matrix}
\usepackage{float}
\floatstyle{boxed} 
\restylefloat{figure}
\usepackage{ytableau}				
\usepackage[numbers]{natbib}
\usepackage{amsmath}
\usepackage{cleveref}
\crefname{figure}{Fig.}{Fig.}
\crefname{section}{\S}{\S\S}
\Crefname{section}{\S}{\S\S}	
\crefformat{section}{\S#2#1#3}

\renewcommand{\thefootnote}{\fnsymbol{footnote}}  

\newcommand{\eps}{\epsilon}

\begin{document}
\setlength{\textheight}{8.0truein}    

\runninghead{A Practical Quantum Algorithm for the Schur Transform}
            {William M. Kirby and Frederick W. Strauch}

\normalsize\textlineskip
\thispagestyle{empty}
\setcounter{page}{1}

\vspace*{0.88truein}

\alphfootnote

\fpage{1}

\centerline{\bf
A PRACTICAL QUANTUM ALGORITHM FOR THE SCHUR TRANSFORM}
\vspace*{0.37truein}
\centerline{\footnotesize
WILLIAM M. KIRBY\footnote{wmk1@williams.edu}}
\vspace*{0.015truein}
\centerline{\footnotesize\it Physics Department, Williams College}
\baselineskip=10pt
\centerline{\footnotesize\it Williamstown, MA 01267,
USA}
\vspace*{10pt}
\centerline{\footnotesize 
FREDERICK W. STRAUCH\footnote{fws1@williams.edu}}
\vspace*{0.015truein}
\centerline{\footnotesize\it Physics Department, Williams College}
\baselineskip=10pt
\centerline{\footnotesize\it Williamstown, MA 01267,
USA}
\vspace*{0.225truein}

\vspace*{0.21truein}

\abstracts{
We describe an efficient quantum algorithm for the quantum Schur transform. The Schur transform is an operation on a quantum computer that maps the standard computational basis to a basis composed of irreducible representations of the unitary and symmetric groups. We simplify and extend the algorithm of Bacon, Chuang, and Harrow, and provide a new practical construction as well as sharp theoretical and practical analyses. Our algorithm decomposes the Schur transform on $n$ qubits into $O\left(n^4\log\left(\frac{n}{\eps}\right)\right)$ operators in the Clifford+T fault-tolerant gate set and uses exactly $2\lfloor\log_2(n)\rfloor-1$ ancillary qubits. We extend our qubit algorithm to decompose the Schur transform on $n$ qudits of dimension $d$ into $O\left(n^{d^2+2}\log^p\left(\frac{n^{d^2+1}}{\eps}\right)\right)$ primitive operators from any universal gate set, for $p\approx3.97$.
}{}{}

\vspace*{10pt}

\vspace*{1pt}\textlineskip    

\textbf{Erratum: previous versions of this paper contained an incorrect analysis of the prior work~\cite{bacon06}. The scaling of the qubit algorithm presented in~\cite{bacon06}, as decomposed into the Clifford+T gateset, is $O(n\,\rm{polylog}(n, 1/\epsilon))$.}

\textbf{Second, the ``Schur transform'' is described in this paper as a passive transformation, i.e., a transformation from a basis in which the computational basis states label local angular momentum numbers to a basis in which the computational basis states label global angular momentum numbers. Hence as an active transformation the algorithm in this paper really implements the inverse Schur transform. Since the algorithm compiles the inverse Schur transform into a quantum circuit, the Schur transform can be obtained by inverting that circuit, so the costs of the algorithm are unchanged.}

\section{Introduction}
\setcounter{footnote}{0}
\renewcommand{\thefootnote}{\alph{footnote}}
\label{introsect}
The \emph{Schur transform} is a useful routine in quantum computing. It is a change of basis on a register of qudits, from the computational basis (composed of tensor products of the individual qudits' states) to an alternate basis called the \emph{Schur basis} \cite{bacon07}. For qubits (dimension $d=2$), the Schur basis is composed of eigenvectors of the global spin of the whole register. For qudits, the situation is more complex, but in both cases the Schur basis exhibits global rather than local symmetry. The Schur transform generalizes the more familiar \emph{Clebsch-Gordan} (CG) transform, which performs this operation on two subsystems.

We can describe the Schur basis more rigorously. If $\mathcal{C}^d$ is the Hilbert space for a qudit of dimension $d$, we can write the Hilbert space for a register of $n$ such qudits as $\left(\mathcal{C}^d\right)^{\otimes n}=\underbrace{\mathcal{C}^d\otimes\mathcal{C}^d\otimes\cdots\otimes\mathcal{C}^d}_{\text{$n$ copies}}$. The Schur basis is a decomposition of $\left(\mathcal{C}^d\right)^{\otimes n}$ into irreducible modules (irreps) of the unitary group $\mathcal{U}_d$ and the symmetric group $\mathcal{S}_n$.

In this paper, we describe an efficient and practical quantum algorithm for the Schur transform. For $n$ qubits, our algorithm decomposes the Schur transform into $O\left(n^4\log\left(\frac{n}{\eps}\right)\right)$ Clifford+T operators\footnote{A later work~\cite{wills23} improves the analysis to $O\left(n^3\log\left(\frac{n}{\eps}\right)\right)$.}, which can be implemented fault-tolerantly \cite{gottesman98,bravyi05}. We refer to this number as the \emph{sequence length}: it is the ``quantum runtime" of our algorithm, since each of the operators in the sequence will need to be performed in order. We also extend our algorithm to $n$ qudits of dimension $d$, for which it decomposes the Schur transform into a sequence of $O\left(n^{d^2+2}\log^p\left(\frac{n^{d^2+1}}{\eps}\right)\right)$ primitive operators from any universal gate set (for $p\approx3.97$). Our qubit algorithm employs exactly $2\lfloor\log_2(n)\rfloor-1$ ancillary qubits. This work simplifies and extends the work by Bacon, Chuang, and Harrow (BCH) in \cite{bacon07, bacon06}: in particular, we provide a practical implementation and explicit analysis of a modification of the qubit algorithm \cite{bacon06} using a minimum of ancilla, and our framework can be extended to a qudit algorithm that expands upon arguments in Section V of \cite{bacon07}.

Throughout this paper, we primarily focus on the special case of qubits ($d=2$), both because it is helpful in developing the right intuitions and pictures of the Schur transform, especially for readers with a background in physics, and because ultimately we provide an explicit implementation of our algorithm for the qubit case. The paper is organized as follows: In the remainder of this section, we describe the specific mathematical background for the Schur transform. We assume some general knowledge of the representation theory of Lie groups and the symmetric group; \cite{goodman98} is a good reference for this material. In \cref{cgsect}, we describe the Clebsch-Gordan transform, which we will use recursively to construct the Schur transform. In \cref{qubit}, we describe our qubit algorithm for the Schur transform, and in \cref{analysis}, we analyze its runtime. In \cref{qudit}, we describe how to extend our qubit algorithm to an algorithm for qudits of arbitrary dimension, and provide an analysis for this general case. In \cref{discussionc}, we summarize and compare our work with that of BCH.

\subsection{Mathematical background}
\label{mathbackground}
\noindent
A group $G$ is \emph{reductive} if every (regular) representation of $G$ is either irreducible or completely reducible.
\begin{theorem}[Isotypic decomposition \cite{bacon07}]
Let $G$ be a reductive group. Then for any $G$-module $V$,
\begin{equation}
	\label{isotypicdecomposition}
	V\stackrel{G}{\cong}\bigoplus_{i=1}^kV^{(i)}
\end{equation}
for some irreps $V^{(i)}$.
\proof{
See \cite{goodman98}
}
\end{theorem}
Note that the $V^{(i)}$ need not all be inequivalent; we will discuss multiplicities shortly. One can show that the unitary group $\mathcal{U}_d$, as well as any finite group, is reductive (see \cite{goodman98}). These facts prove the following theorem:
\begin{theorem}
\label{isotypicdecompositionthm}
Any module of the symmetric group $\mathcal{S}_n$ (since it is finite), or the unitary group $\mathcal{U}_d$, can be decomposed into a finite direct sum of irreducible submodules, as in \eqref{isotypicdecomposition}.
\end{theorem}

From \cref{isotypicdecompositionthm} we obtain the Schur transform. We denote the general linear group acting on $(\mathcal{C}^d)^{\otimes n}$ (the Hilbert space of a register of $n$ $d$-dimensional qudits) by $GL\left((\mathcal{C}^d)^{\otimes n}\right)$. To describe the Schur transform, consider the following pair of representations:
\begin{itemize}
	\item$P:S_n\rightarrow GL\left((\mathcal{C}^d)^{\otimes n}\right)$, defined by
		\begin{equation}
			\label{symmetricrep}
			P(s)|\phi_1\rangle\otimes|\phi_2\rangle\otimes\cdots\otimes|\phi_n\rangle=|\phi_{s^{-1}(1)}\rangle\otimes|\phi_{s^{-1}(2)}\rangle\otimes\cdots\otimes|\phi_{s^{-1}(n)}\rangle
		\end{equation}
		where $s$ is any permutation (in $S_n$); that is, $P(s)$ permutes the component states according to $s$.
	\item$Q:\mathcal{U}_d\rightarrow GL\left((\mathcal{C}^d)^{\otimes n}\right)$, defined by
		\begin{equation}
			\label{unitaryrep}
			Q(U)|\phi_1\rangle\otimes|\phi_2\rangle\otimes\cdots\otimes|\phi_n\rangle=U|\phi_1\rangle\otimes U|\phi_2\rangle\otimes\cdots\otimes U|\phi_n\rangle
		\end{equation}
		for any $U\in\mathcal{U}_d$; that is, $Q(U)$ applies $U$ to each qudit.
\end{itemize}

By \cref{isotypicdecompositionthm}, $(\mathcal{C}^d)^{\otimes n}$ can be decomposed into a finite direct sum of irreducible submodules of either of these representations. Let the irreducible $Q$-modules be denoted $\mathcal{Q}_\lambda^d$, for some index $\lambda$, and let the irreducible $P$-modules be denoted $\mathcal{P}_\lambda$. Then we can write
\begin{equation}
	\label{decompunitary}
	(\mathcal{C}^d)^{\otimes n}\stackrel{\mathcal{U}_d}{\cong}\bigoplus_\lambda\mathcal{M}_\lambda\otimes\mathcal{Q}_\lambda^d
\end{equation}
and
\begin{equation}
	\label{decompsymmetric}
	(\mathcal{C}^d)^{\otimes n}\stackrel{\mathcal{S}_n}{\cong}\bigoplus_\lambda\mathcal{N}_\lambda\otimes\mathcal{P}_\lambda
\end{equation}
where $\mathcal{M}_\lambda$ and $\mathcal{N}_\lambda$ are multiplicity spaces.
These expressions can be combined by applying \emph{Schur duality} \cite{bacon07}, which states that the multiplicity spaces $\mathcal{M}_\lambda$ are isomorphic to the irreps $\mathcal{P}_\lambda$ (and equivalently, the multiplicity spaces $\mathcal{N}_\lambda$ are isomorphic to the irreps $\mathcal{Q}_\lambda^d$).
Therefore, we can consolidate our two expressions \eqref{decompunitary} and \eqref{decompsymmetric} into one:
\begin{equation}
	\label{schurdecompmodules}
	\left(\mathcal{C}^d\right)^{\otimes n}\stackrel{\mathcal{U}_d\times S_n}{\cong}\bigoplus_\lambda\mathcal{P}_\lambda\otimes\mathcal{Q}_\lambda^d
\end{equation}
A derivation of this result can be found in \cite{bacon07}.

We have not yet stated what the index $\lambda$ is, nor how to find the irreps $\mathcal{P}_\lambda$ and $\mathcal{Q}_\lambda^d$. We now give these results, refering the reader to other sources for their derivations.

The index $\lambda$ runs over all \emph{partitions} of $n$ \cite{bacon07}. For $\lambda$ to partition $n$ means that $\lambda=(\lambda_1,\lambda_2,...,\lambda_\ell)$ for positive integers $\lambda_i$ such that $\sum_{i=1}^\ell\lambda_i=n$ and $\lambda_1\ge\lambda_2\ge\cdots\ge\lambda_\ell>0$. We call $\ell$ the \emph{degree} of $\lambda$, and we write $\lambda\vdash n$. There is one distinct irrep of $\mathcal{U}_d$ and one distinct irrep of $\mathcal{S}_n$ for each $\lambda\vdash n$.

To obtain the dimensions of the irreps, we use Young diagrams and tableaux. The \emph{Young diagram} associated to a partition $\lambda$ is an array of boxes, with $\lambda_i$ boxes in the $i$th row. For example, if $\lambda=(3,1)$, then the Young diagram of $\lambda$ is
\begin{equation}
	\begin{ytableau}\ &\ &\ \\\ \end{ytableau}
\end{equation}
A \emph{tableau} (plural: \emph{tableaux}) is an assignment of integers to the boxes in a Young diagram. We call a tableau associated to some partition $\lambda$ a $\lambda$-tableau. Here are a few $(3,1)$-tableaux
\begin{equation}
	\label{tableauxexamples}
	\begin{ytableau}1&2&3\\4\end{ytableau}\ ,\quad\begin{ytableau}1&1&1\\2\end{ytableau}\ ,\quad\begin{ytableau}4&4&3\\1\end{ytableau}
\end{equation}
A \emph{standard tableau} is a tableau whose entries are strictly increasing across the rows and down the columns: for example, the first tableau in \eqref{tableauxexamples} is a standard $(3,1)$-tableau; we further restrict the entries to be the set $\{1,2,...,n\}$. This last restriction is not part of the usual definition of a standard tableau, but it will simplify our usage. A \emph{semistandard tableau} is a tableau whose entries are weakly increasing across the rows and strictly increasing down the columns: for example, the first and second tableaux in \eqref{tableauxexamples} are semistandard $(3,1)$-tableaux.

We find the dimensions of our irreps by counting $\lambda$-tableaux:
\begin{align}
	&\text{dim}\left(\mathcal{Q}_\lambda^d\right)=\text{\# of semistandard $\lambda$-tableaux with entries in $\{1,2,...,d\}$}\label{dimq}\\
	&\text{dim}\left(\mathcal{P}_\lambda\right)=\text{\# of standard $\lambda$-tableaux}\label{dimp}
\end{align}
\eqref{dimq} has the following useful corollary:
\begin{corollary}
\label{deglemma}
For $\lambda\vdash n$ (for some integer $n$), if $\text{deg}(\lambda)>d$, then $dim\left(\mathcal{Q}_\lambda^d\right)=0$.
\end{corollary}

Our critical result in this section was \eqref{schurdecompmodules}. We define the \emph{Schur basis} to be a basis in which \eqref{schurdecompmodules} is an equality. The \emph{Schur transform} is a change of basis from the computational basis to the Schur basis. Since we know from \eqref{schurdecompmodules} that the $\mathcal{S}_n$-irreps $\mathcal{P}_\lambda$ act as the multiplicity spaces for the $\mathcal{U}_d$-irreps $\mathcal{Q}_\lambda^d$ and vice versa, we know that whether we structure our implementation in terms of \eqref{decompunitary} or \eqref{decompsymmetric}, we will always obtain \eqref{schurdecompmodules}. Through most of this document, we will work in terms of \eqref{decompunitary}. Thus, since the irreps $\mathcal{P}_\lambda$ are the multiplicity spaces for the irreps $\mathcal{Q}_\lambda^d$ in \eqref{schurdecompmodules}, by \eqref{dimp} we have
\begin{equation}
	\label{dimseq}
	\text{multiplicity of }\mathcal{Q}_\lambda^d=\text{\# of standard $\lambda$-tableaux}
\end{equation}

\subsection{Applications and previous work}
\noindent
A few of the better known applications of the Schur transform are decoherence-free encoding \cite{zanardi97,lidar98,bishop09}, quantum hypothesis testing \cite{audenaert08,hayashi02_2}, spectrum estimation \cite{keyl01,gill02}, entanglement concentration \cite{hayashi02,blumekohout09}, and reference-frame independent quantum communication \cite{bartlett03}.

David Bacon, Isaac Chuang, and Aram Harrow published a pair of papers \cite{bacon07,bacon06} in which they first described efficient quantum algorithms for the Schur transform.  Their work provides an analysis of the circuit and its runtime, showing that the runtime is linear in $n$ up to logarithmic factors.  A subsequent paper \cite{berg12} describes a generalization of BCH's construction.

Our algorithm is inspired by the work of BCH and shares its outermost layer of structure with their construction: it recursively decomposes the Schur transform into a succession of operators built out of Clebsch-Gordan (CG) transforms (see Fig.\ 1). Our algorithm differs from that of BCH in the implementation of each individual recursion step. The insight that motivates our implementation of the recursion step goes roughly as follows:

\begin{figure}[h!]
	\caption{Algorithm structure. Ancillary qudits are not shown explicitly. The total amount of space added in the ``Add space from ancillas" step is logarithmic in the total dimension of the Hilbert space.}
	\centering
	\includegraphics[width=5.5in]{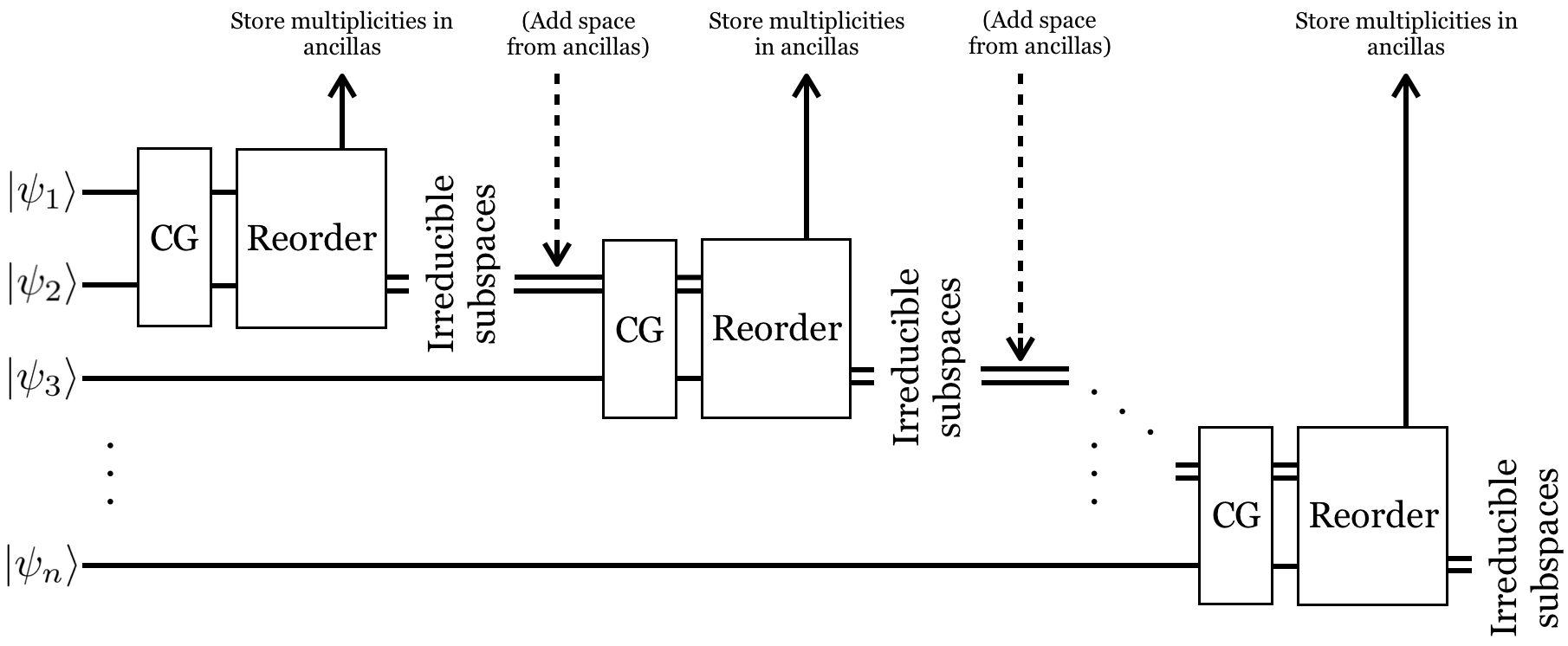}
\end{figure}

At each large-scale step in the algorithm, we have a decomposition of the full Hilbert space into some set of subspaces, each of which has an irrep in the decomposition \eqref{schurdecompmodules} for some subset of the qudits in the register. We will show herein that a careful reordering of these subspaces at each step simplifies the algorithm in such a way that an explicit analysis is possible. The reordering also makes the structure of the algorithm directly reflect the structure of the underlying mathematical objects. To perform the reordering, we add a logarithmic number of ancillary qudits to the register. Then, roughly speaking, the reordering arranges the irreducible representations of the globally-symmetric unitary group (for the current iteration) over the states of the ancillary qudits, which then act as indices. This allows the next recursion step to act in parallel on each of the distinct irreducible representations, whose individual dimensions are logarithmic in the total dimension, which in turn allows the algorithm to achieve polynomial instead of exponential runtime.  This way of structuring of the algorithm is one of the main contributions of our work, as it allows for a relatively simple way to encode the irrep labels with a minimal number of ancilla.  We will discuss this approach in detail in \cref{qubit} (for qubits) and \cref{qudit} (for qudits).

\section{The Clebsch-Gordan Transform}
\label{cgsect}
In this section we describe the \emph{Clebsch-Gordan} (CG) transform, which will form the recursion step in our construction of the Schur transform in the next section.

\subsection{Mathematical perspective}
\label{cgmath}

For $\mu\vdash n$ and $\nu\vdash m$, let $\mathcal{Q}_\mu^d$ and $\mathcal{Q}_\nu^d$ be irreps of $\mathcal{U}_d$ acting on $n$ and $m$ qudits. By \cref{isotypicdecompositionthm}, we can decompose the tensor product module into irreps:
\begin{equation}
	\label{cg1}
	\mathcal{Q}_\mu^d\otimes\mathcal{Q}_\nu^d\stackrel{\mathcal{U}_d}{\cong}\bigoplus_{\lambda\vdash(n+m)}\mathcal{M}_\lambda\otimes\mathcal{Q}_\lambda^d
\end{equation}
where $\mathcal{M}_\lambda$ is the multiplicity space associated to the irrep $\mathcal{Q}_\lambda^d$. The CG transform maps the basis on the left-hand side of \eqref{cg1} to the basis on the right-hand side. In the case $\nu=(1)$ (which will turn out to be the case we are interested in), \eqref{cg1} simplifies to the following expression:
\begin{equation}
	\label{cg2}
	\mathcal{Q}_\mu^d\otimes\mathcal{Q}_{(1)}^d\stackrel{\mathcal{U}_d}{\cong}\bigoplus_\lambda\mathcal{Q}_\lambda^d
\end{equation}
where $\lambda=\mu+\textbf{e}_j$ for some $j$ such that $\lambda$ is a valid partition and has degree less than or equal to $d$ \cite{bacon07}. We can visualize this schematically in terms of Young diagrams: for example, if $d=2$ and $\mu=(3,1)$, then the corresponding Young diagram is $\begin{ytableau}\ &\ &\ \\\ \end{ytableau}$, and \eqref{cg2} can be visualized as
\vspace{+10pt}
\begin{equation}
	\begin{ytableau}\ &\ &\ \\\ \end{ytableau}\otimes\begin{ytableau}\ \end{ytableau}\quad\stackrel{\mathcal{U}_d}{\cong}\quad\begin{ytableau}\ &\ &\ &\ \\\ \end{ytableau}\oplus\begin{ytableau}\ &\ &\ \\\ &\ \end{ytableau}
\end{equation}
If $d$ were greater than $2$, the Young diagram $$\begin{ytableau}\ &\ &\ \\\ \\\ \end{ytableau}$$ would also appear on the right-hand side. Thus we can think of the version of the CG transform that we are interested in as the ``add-a-box" operation, which maps the standard tensor product basis for $\mathcal{Q}_\mu^d\otimes\mathcal{Q}_{(1)}^d$ to a basis in which \eqref{cg2} is an equality \cite{bacon07}.

\subsection{Physical perspective}
\label{cgphys}
We can also present the CG transform for $d=2$ in a context more familiar to physicists, most easily described in terms of spin.

\begin{definition}
The \emph{total spin operator} (squared) $\textbf{J}^2$ is defined by
\begin{equation}
	\textbf{J}^2|j,m\rangle=j(j+1)\hbar^2|j,m\rangle
\end{equation}
\end{definition}
\begin{definition}
The \emph{spin projection operator} $J^z$ is defined by
\begin{equation}
	J^z|j,m\rangle=m\hbar|j,m\rangle
\end{equation}
\end{definition}
A particular spin system is defined by a particular value of $j$: for example, spin-$1/2$ refers to $j=1/2$. A spin-$j$ system has a basis made up of the eigenvectors of the spin-projection operator, which are indexed by the values of $m$. Thus, a spin-$j$ system is $(2j+1)$-dimensional \cite{townsend12}.

The computational basis is labeled by the spins of the subsystems, which makes it convenient if we want to operate on the subsystems independently. However, the composite system has its own total spin and spin-projection, so we can transform to a basis parametrized by these quantities instead. If $\textbf{J}^2_1$ and $\textbf{J}^2_2$ are the total spin operators for the two subsystems, and $J^z_1$ and $J^z_2$ are the spin-projection operators, then the composite total spin and spin-projection operators are
\begin{equation}
	\textbf{J}^2=(\textbf{J}_1+\textbf{J}_2)^2,\quad\quad\quad J^z=J^z_1+J^z_2
\end{equation}
\cite{sakurai85}. One can show \cite{sakurai85} that for a system composed of a spin-$j_1$ and a spin-$j_2$, the composite total spin has eigenvalues
\begin{equation}
	j=j_1+j_2,j_1+j_2-1,...,|j_1-j_2|
\end{equation}
with the corresponding possible values of $m$.

We are now in a position to define the CG transform (for $d=2$) from a physics perspective:
\begin{definition}
The \emph{Clebsch-Gordan transform} on a system of two spins maps the computational basis to the basis composed of eigenvectors of the composite spin operators.
\end{definition}
There are a number of ways to calculate the CG transform classically (for example, \cite{schertler96} and \cite{burke10}). A simple method is to begin with the (single) computational basis state with the highest value of the composite total spin and spin projection, and apply the composite lowering operator repeatedly to obtain all of the states with that total spin, then select the next highest composite total spin and spin projection, and so forth.

For $d>2$, generalizations of the above calculation exist (see \cite{alex11} for a nice example). For our present purpose it is sufficient to know that the CG transforms for arbitrary $d$ are classically and efficiently calculable, since in our algorithm that part of the computation will be classical.

\subsection{Putting the pieces together}
\label{puttingpieces}
We would like to reconcile the two descriptions we have given for the CG transform in the last two sections, starting with the case $d=2$. We begin by noting the immediate similarities:
\begin{enumerate}
	\item In the mathematical description of the CG transform, we specialized to decomposing a tensor product of $\mathcal{Q}_\mu^d$ (for some $\mu\vdash n$) and $\mathcal{Q}_{(1)}^d$ into irreps. We can think of each of the irreps $\mathcal{Q}_\mu^d$ and $\mathcal{Q}_{(1)}^d$ as subsystems to be combined into a larger composite system.
	
	\item The physical description of the CG transform is also based on the concept of combining two subsystems into a larger composite system, with the only qualitative differences being that both of the subsystems are assumed to be definite spins, and that the spins are allowed to be different (recall that in the mathematical description we assumed that all of the boxes had entries in $1,2,...,d$ for the semistandard tableaux).
	
\end{enumerate}

As it turns out, the generality of the physical CG transform in the dimensions of the subsystems and the generality of the mathematical CG transform in the first irrep in the tensor product are intimately related. This is a consequence of the powerful fact that from a quantum informational perspective, two systems with the same dimension are equivalent as long as they carry the same representation of $\mathcal{U}_d$ (in the current case, $\mathcal{U}_2$). We take advantage of this fact whenever we talk about a qudit in the abstract: the structure is the same whether the physical qudit is an ion, a photon, or a superconducting quantum circuit. In the $d=2$ case, the equivalence is between an irrep (in the mathematical picture) and a spin (in the physical picture). Since both are perfectly legitimate quantum systems in their own rights, as long as they have the same dimensions and carry the same representation of $\mathcal{U}_2$ they can be treated as equivalent.

The operation we will want to perform with our CG transforms is the addition of a single qudit to a larger register of qudits, which is why we constrained the second irrep in our mathematical description to be $\mathcal{Q}_{(1)}^d$. The other qudits in the register will have previously been decomposed into their irreps by a Schur transform, so we can consider their irreps separately: hence the generality of the partition $\mu$ that labels the first irrep in our mathematical description \eqref{cg2}.
Any particular $\mathcal{Q}_\mu^d$ is just a Hilbert space whose dimension is determined by the number of semistandard $\mu$-tableaux with entries in $1,2,...,d$ (as given in \eqref{dimq}).
This is why, for example, it is correct to say that a composite system of two spin-$1/2$ particles decomposes into a spin-$1$ and a spin-$0$: the three spin-$1$ states correspond to the irrep $\mathcal{Q}_{(2)}^2$, and the singlet state, with spin-$0$, corresponds to the irrep $\mathcal{Q}_{(1,1)}^2$ (see Fig.\ 2).
\begin{figure}[h]
\caption{CG (Schur) transform on two qubits}
\centering
\begin{description}
	\item[Triplets (spin-$1$) $=$ $\mathcal{Q}_{(2)}^2$:]
		$$\begin{Bmatrix}
			|m=1\rangle=|\frac{1}{2},\frac{1}{2}\rangle\\
			|m=0\rangle=\frac{1}{\sqrt{2}}\left(|\frac{1}{2},-\frac{1}{2}\rangle+|-\frac{1}{2},\frac{1}{2}\rangle\right)\\
			|m=-1\rangle=|-\frac{1}{2},-\frac{1}{2}\rangle
		\end{Bmatrix}
		\quad\quad\Longleftrightarrow\quad\quad
		\begin{Bmatrix}
			\begin{ytableau}1&1\end{ytableau}\\
			\begin{ytableau}1&2\end{ytableau}\\
			\begin{ytableau}2&2\end{ytableau}
		\end{Bmatrix}$$
	\item[Singlet (spin-$0$) $=$ $\mathcal{Q}_{(1,1)}^2$:]
		$$\begin{Bmatrix}
			|m=0\rangle=\frac{1}{\sqrt{2}}\left(|\frac{1}{2},-\frac{1}{2}\rangle-|-\frac{1}{2},\frac{1}{2}\rangle\right)
		\end{Bmatrix}
		\quad\quad\Longleftrightarrow\quad\quad
		\begin{Bmatrix}
			\begin{ytableau}1\\2\end{ytableau}
		\end{Bmatrix}$$
\end{description}
\end{figure}

We should not interpret the correspondences in Fig.\ 2 as direct identifications between the state vectors and tableaux, but the spanned subspaces are correct. The main point is that we can now perform, for example, the CG transform on $\mathcal{Q}_{(2)}^2\otimes\mathcal{Q}_{(1)}^2$ (i.e., add another qubit), by treating the vector space $\mathcal{Q}_{(2)}^2$ as if it were a spin-$1$ particle and using the physicists' CG transform.
Thus, as we will discuss in the next section, we can use a direct sum of physicists' CG transforms, each of which acts on two systems, to perform a ``super-CG transform" that adds one qubit to a register of qubits that have already been decomposed into irreps: this will be the recursion step in our implementation of the Schur transform.

\section{Implementation for qubits}
\label{qubit}
In this section we will describe a quantum algorithm that performs the Schur transform on $n$ qubits. We will then prove that this algorithm is efficient: in particular, we will show that it decomposes the Schur transform on $n$ qubits into a sequence of
\begin{equation}
	O(n^3)
\end{equation}
two-level gates. In fact, the number of two-level gates is bounded by $2n^3$; an even tighter upper bound is given in \eqref{runtime}. A two-level gate is unitary operator that only acts on two dimensions, i.e., is isomorphic to a $2\times2$ unitary. Each two-level gate can be decomposed to accuracy $\eps$ using known methods \cite{kliuchnikov13,kliuchnikov13b,selinger14} into $O(n\log(1/\eps))$ gates from the Clifford+T set, which can be implemented fault-tolerantly \cite{gottesman98,bravyi05}.

\subsection{Recursion structure}
Our algorithm is recursive in its outermost layer of structure: this is the element that is shared with BCH's construction \cite{bacon07}. The iteration step looks like this:
\begin{itemize}
	\item Suppose you are given a basis for the Hilbert space of $k$ qubits that is composed of eigenvectors of the total spin and spin projection operators for the \emph{whole} system. (We refer to these operators as the \emph{global spin operators}.)
	\item Then the basis can be broken up into a disjoint union of the bases for several subsystems, each of which has a definite value for the global total spin. For example, if we have $k=3$ qubits, the Hilbert space decomposes into a spin-$3/2$ subspace and two spin-$1/2$ subspaces.
	\item In other words, assume that we are already in the Schur basis for $k$ qubits.
	\item Then in order to lift our Schur basis for $k$ qubits to the Schur basis for $k+1$ qubits, all we need to do is apply the appropriate CG transform to each spin-subspace. For example, if we begin with $k=3$ qubits, to add a fourth qubit, we apply the CG transform that adds a spin-$1/2$ to a spin-$3/2$ to the first subspace, which outputs a spin-$2$ subspace and a spin-$1$ subspace. We also apply the CG transform that adds a spin-$1/2$ to a spin-$1/2$ to each of the spin-$1/2$ subspaces we already have: each of these outputs a spin-$1$ subspace and a spin-$0$ subspace. So in total we end up with a spin-$2$ subspace, $3$ spin-$1$ subspaces, and $2$ spin-$0$ subspaces: the Schur basis for $4$ qubits.
	\item Thus we can iterate from the Schur basis for $k$ qubits to the Schur basis for $k+1$ qubits by applying something like a direct sum of CG transforms: see Fig.\ 3. Continuing the above example, the red block in Fig.\ 3 corresponds to the CG transform that adds a spin-$1/2$ to a spin-$3/2$, and the two blue blocks each correspond to the CG transform that adds a spin-$1/2$ to a spin-$1/2$. Fig.\ 3 does \emph{not} represent the final form of the iteration operator, which we will describe in the next section, but it is a good picture to start from.
		\begin{figure}[h!]
			\caption{Direct sum of CG transforms to lift the Schur basis for $3$ qubits to the Schur basis for $4$ qubits.}
			\centering
			\includegraphics[width=2in]{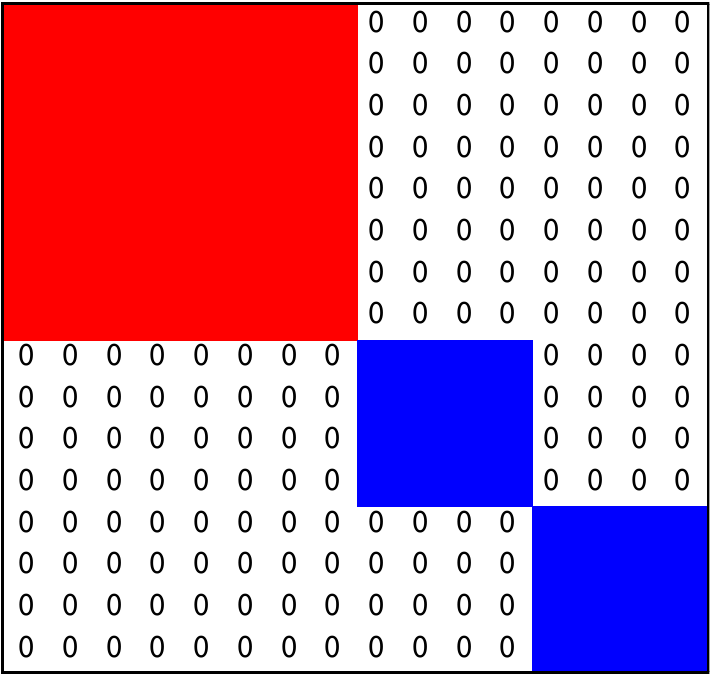}
		\end{figure}
	\item Since the Schur basis for $1$ qubit is identical to the computational basis, we can repeatedly apply our iteration operator to obtain the Schur transform on any number of qubits.
	\item The key to our algorithm is structuring the iteration operator so that it can be implemented efficiently.
\end{itemize}

\subsection{Ordering the basis}
\label{ordering}
We make careful choices of orderings for the bases that appear at each step in our algorithm. By adding a logarithmic (in $n$) number of ancillary qubits, we can index the various levels of structure in the Schur basis: the multiplicities of the irreps (spin-subspaces), the inequivalent irreps themselves, and finally the states within the irreps. The ordering we will construct is motivated by the observation that most of the inequivalent irreps in the Schur basis have multiplicities greater than one. Therefore, when we perform the iteration step described above, all the copies of each distinct irrep will have a new qubit added to them via the same CG transform. Our ordering allows us to implement each of these distinct CG transforms only once, in parallel over the irreps to which they need to be applied.

We will begin by describing in general the ordering of the Schur basis at any step in the iteration. This will make our description of the iteration step itself straightforward. After any step in the iteration, the Schur basis is encoded in a larger number of qubits than $n$, so there will be ``ghost" states that are not used in the encoding. The encoding qubits are divided into three categories; during the iteration step these categories will become fluid, but between steps they are well-defined. The categories are called:
\begin{enumerate}
	\item\textbf{seq}: qubits that encode the multiplicities of each irrep
	\item\textbf{par}: qubits that label the inequivalent irreps
	\item\textbf{stat}: qubits that encode the states within the irreps
\end{enumerate}
We will generally adhere to this order when thinking about tensor products of the qubits. Thus, if we think of the \textbf{seq}, \textbf{par}, and \textbf{stat} qubits as subsystems with some dimensions determined by the numbers of qubits in each, we obtain vectors with the following form:
\begin{equation}
	\label{encodingform}
	|\textbf{seq}\rangle\otimes|\textbf{par}\rangle\otimes|\textbf{stat}\rangle
\end{equation}
We note that a similar analysis of these registers, as used in the BCH algorithm, was performed in \cite{blumekohout09}.

A particular state in the Schur basis is encoded in the following way: if the state is the $k$th state in the $i$th copy of the $j$th inequivalent irrep, then in the form \eqref{encodingform}, it is labeled $|i\rangle\otimes|j\rangle\otimes|k\rangle$. For example, suppose $n=3$ at the current step, and suppose we want to label the second state in the first spin-$1/2$ subspace. Indexing from $1$, ``second state" translates to $\textbf{stat}=2$, ``spin-$1/2$ subspace" translates to $\textbf{par}=2$ (assuming we put the spin-$3/2$ subspace first in \textbf{par}), and ``\emph{first} (spin-$1/2$ subspace)" translates to $\textbf{seq}=1$. Thus this state is labeled $|1\rangle\otimes|2\rangle\otimes|2\rangle$.

One complication is that within \textbf{seq}, the encodings of the multiplicities will not always be ordered in the most intuitive way; but they will be ordered in a calculable way. We will see shortly that this is a consequence of the structure of the iteration step.  Before we discuss the structure of the iteration, though, let us consider an example of the basis ordering.

\noindent\textbf{Example:}

Suppose $n=4$. Then we have three inequivalent irreps, corresponding to the three partitions $\lambda\vdash4$ with degree 2: $\lambda=(4,0)$, $\lambda=(3,1)$, and $\lambda=(2,2)$ (see \cref{deglemma}; here we relax our definition of degree-2 to include $(4,0)$ as a degree-2 partition). To get the multiplicities and dimensions of the irreps, we use \eqref{dimq} and \eqref{dimseq}. The dimensions \eqref{dimq} are easy to evaluate: the number of degree-$2$ semistandard tableaux with entries in $\{1,2\}$ is given by the number of possible locations for the first $2$ in the first row, which is $\lambda_1-\lambda_2+1=n+1-2\lambda_2$. (The multiplicities \eqref{dimseq} can be calculated by the hook formula \cite{sagan01}, although this is not necessary for the actual running of our algorithm.) We obtain:
\begin{align}
	&\mathcal{Q}_{(4,0)}^2:\quad\text{multiplicity}=1,\quad\text{dimension}=5\nonumber\\
	&\mathcal{Q}_{(3,1)}^2:\quad\text{multiplicity}=3,\quad\text{dimension}=3\nonumber\\
	&\mathcal{Q}_{(2,2)}^2:\quad\text{multiplicity}=2,\quad\text{dimension}=1
\end{align}
(Physically, these irreps correspond to spin-$2$, spin-$1$, and spin-$0$ subspaces, respectively.) We can represent one possible ordering for these irreps schematically by
\begin{equation}
	\label{inputvector}
	\centering{\vcenter{\includegraphics[height=2.5in]{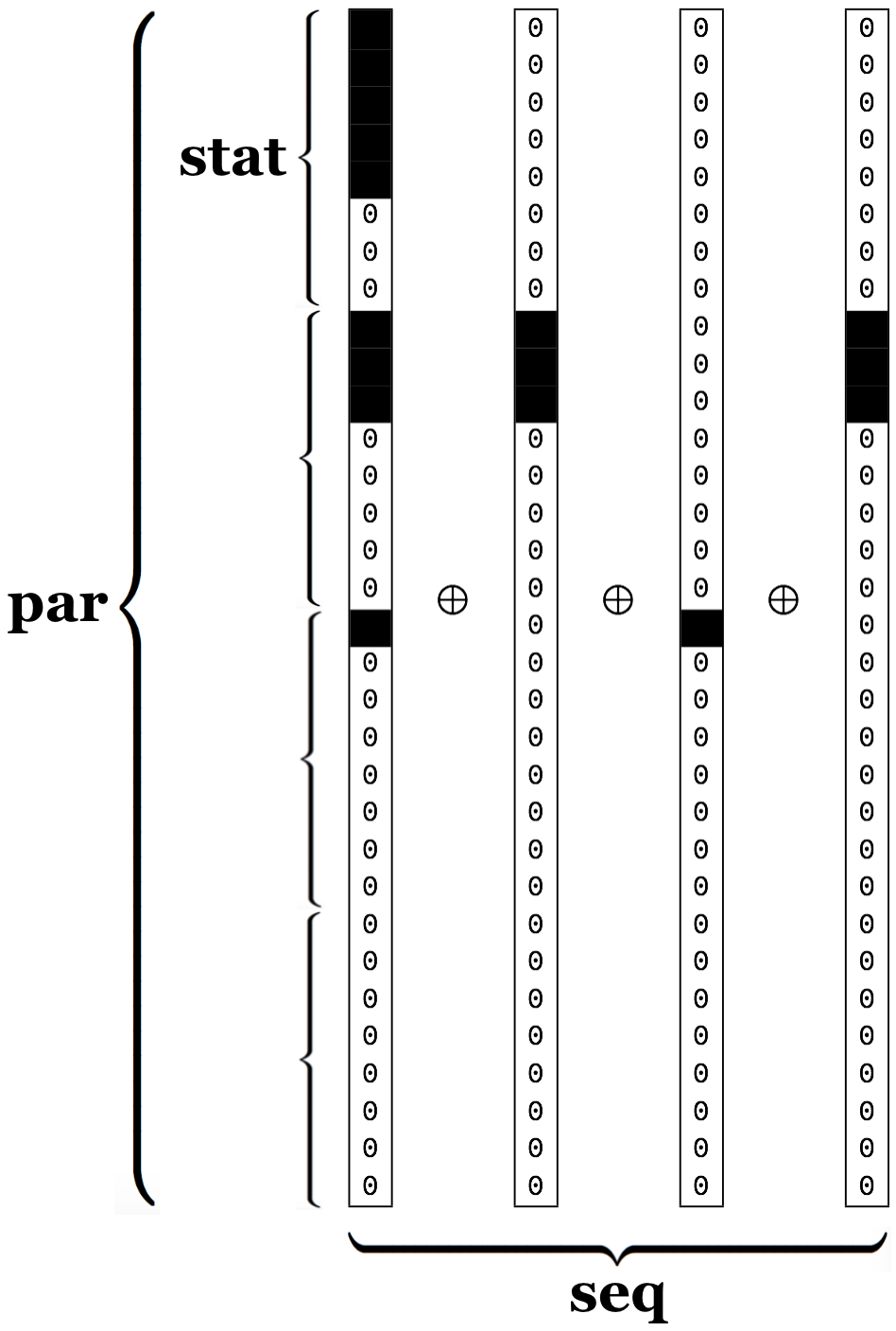}}}
\end{equation}
Here the solid black entries mark the locations of the encoded irreps, and the zeroes mark ghost entries that are not used in the encoding. The largest dimension is $5$, so we need $3$ qubits ($8$ states) in \textbf{stat} to encode the states within the irreps. There are $3$ inequivalent irreps, so we need $2$ qubits ($4$ states) in \textbf{par} to identify the inequivalent irreps. The largest multiplicity is $3$, so we need $2$ qubits ($4$ states) in \textbf{seq} to index the copies of the irreps. Thus in \eqref{inputvector}, the states of \textbf{seq} index the columns. Within each column, there are four ``slots," each comprising $8$ states. The slots are indexed by the states of \textbf{par}, and each inequivalent irrep is assigned to a particular slot: for example, the copies of $\mathcal{Q}_{(3,1)}^2$ ($3$-dimensional) all appear in the second slot in their column. The states of \textbf{stat} index the states of the irreps within each slot. We will continue to use the term \emph{slot} to refer to the set of states used to encode a particular irrep, and the term \emph{column} to refer to a set of slots used to encode all of the inequivalent irreps.

So, for example, to find the second state in the second copy of $\mathcal{Q}_{(3,1)}^2$, we first find the appropriate column for the second copy of $\mathcal{Q}_{(3,1)}^2$. We then find the second slot within this column, since $\mathcal{Q}_{(3,1)}^2$ is the second inequivalent irrep: in this case, the second slot is entries $9$ through $16$ (indexing from $1$). We then find the second state within this slot. So in \eqref{inputvector}, the second state in the second copy of $\mathcal{Q}_{(3,1)}^2$ is encoded in the $10$th entry of the second column, or in the notation of \eqref{encodingform}, $|2\rangle\otimes|2\rangle\otimes|2\rangle$.

\noindent\rule{3in}{0.4pt}

The advantage of the ordering we have just described is that the iteration step has the same action on each of the states of \textbf{seq} (the columns in \eqref{inputvector}). Thus the complexity of the iteration step will be independent of the dimension of \textbf{seq}, which is determined by the multiplicities of the irreps. We will see that the exponential growth (with $n$) in the dimension of the overall Hilbert space is absorbed into the dimension of \textbf{seq}, with the dimensions of \textbf{par} and \textbf{stat} growing only as polynomials in $n$, so since the complexity of the iteration step is independent of the dimension of \textbf{seq}, we obtain polynomial complexity in $n$ for our algorithm.

\subsection{Iteration step}
\label{iterationstep}
We now describe the iteration step that takes a Schur basis of $n$ qubits (as described in the previous section) and lifts it to a Schur basis of $n+1$ qubits. The iteration step has three pieces:
\begin{enumerate}
	\item Add the new computational qubit, as well as ancillary qubits if necessary.
	\item Perform CG transforms to lift to the new Schur basis. We refer to this step as the ``super-CG transform," since it is a direct sum of CG transforms.
	\item Reorder according to the new Schur basis. We refer to this step as the ``reordering transform."
\end{enumerate}
Let us expand these substeps:

\noindent\textbf{1.}
Add the new computational qubit to the bottom of the register (the end of the tensor product). We can think of adding the new computational qubit as doubling the dimensions of the irrep slots discussed in the previous section. We will see in step 3 that whenever
\begin{equation}
	\label{changepar}
	\lfloor\log_2(n+1)\rfloor\neq\lfloor\log_2(n)\rfloor
\end{equation}
we have to add an additional pair of ancillary qubits in order to properly reorder the Schur basis. Since this happens only ``logarithmically often" in $n$, the total number of ancillary qubits thus added will be logarithmic in $n$.
	
\noindent\textbf{2.}
As we discussed in \cref{ordering}, each state of \textbf{seq} labels a column composed of slots corresponding to each of the inequivalent irreps $\mathcal{Q}_\mu^2$. A CG transform is associated to each slot: the CG transform decomposes the composite system of $\mathcal{Q}_\mu^2$ and a new qubit (that is, $\mathcal{Q}_\mu^2\otimes\mathcal{Q}_{(1)}^2$) into a direct sum of irreps, as given in \eqref{cg2}. So the iteration step acts identically on each column, and its action is a direct sum of CG transforms.
	
\noindent\textbf{3.}
After performing the CG transforms, the basis is composed of irreps for $n+1$ qubits, but it is not yet ordered according to the scheme described in \cref{ordering}, so the final step is to perform the reordering. The nature of the reordering depends on whether \eqref{changepar} holds. This condition comes from counting the number of qubits required to encode \textbf{par} and \textbf{stat}:
		
The number of qubits in \textbf{par} is determined by the number of inequivalent irreps, which is equal to the number of degree-$2$ partitions of $n$ (hence the name ``par," for ``partition"). For $\lambda=(\lambda_1,\lambda_2)\vdash n$, $\lambda_2$ can take any value from $0$ to $\lfloor n/2\rfloor$, so the number of degree-$2$ partitions of $n$ is $\lfloor n/2\rfloor+1$. Thus the number of qubits in \textbf{par} is
\begin{equation}
	\label{npar}
	|\text{\textbf{par}}|=\lceil\log_2\left(\lfloor n/2\rfloor+1\right)\rceil
\end{equation}
When $n\rightarrow n+1$, the value of \eqref{npar} increases by $1$ if and only if \eqref{changepar} is true.
		
The number of qubits in \textbf{stat} is determined by the highest dimension of any irrep (``stat" stands for ``state"). These dimensions are given by \eqref{dimq}, so prior to adding the new qubit, the largest such dimension is associated to the partition $(n,0)$: $\mathcal{Q}_{(n,0)}^2$ has dimension $n+1$. Thus the number of qubits in \textbf{stat} is
\begin{equation}
	\label{nstat}
	|\text{\textbf{stat}}|=\lceil\log_2(n+1)\rceil
\end{equation}
When $n\rightarrow n+1$, the value of \eqref{nstat} also increases by $1$ if and only if \eqref{changepar} is true.
		
The number of qubits in \textbf{seq} must be sufficient to encode the highest multiplicity of any irrep. The multiplicity of $\mathcal{Q}_\lambda^2$ is given by the number of standard $\lambda$-tableaux. This number can be calculated exactly, but we will use a simpler approximation. We can imagine building any standard $\lambda$-tableau by inserting the integers $1$ to $n$ one at a time in order into the boxes in the Young diagram of $\lambda$. Each insertion of an integer must be into the leftmost open box in one of the two rows in the Young diagram, so we have at most two choices for where to place each integer. Further, when we insert $1$ we have no choice, since it must be in the upper left box in the diagram, and when we insert $n$ we have no choice, since it must fill the one remaining open box in the diagram. So for \emph{any} $\lambda\vdash n$, we make a sequence of at most $n-2$ binary choices to generate any standard $\lambda$-tableau (hence the name ``seq," for ``sequence"), and thus the number of standard $\lambda$-tableaux is bounded by $2^{n-2}$. Therefore, we take the number of qubits in \textbf{seq} to be
\begin{equation}
	\label{nseq}
	|\text{\textbf{seq}}|=\log_2\left(2^{n-2}\right)=n-2
\end{equation}
Since the total number of encoding qubits must be at least $n$, at least one of $|\text{\textbf{seq}}|$, $|\text{\textbf{par}}|$, and $|\text{\textbf{stat}}|$ must scale linearly with $n$, so we know that \eqref{nseq} is not a drastic overestimate.

Thus, we see that at each iteration, the new computational qubit ends up as a \textbf{seq} qubit. Every time \eqref{changepar} is true, we increase $|\text{\textbf{par}}|$ and $|\text{\textbf{stat}}|$ by $1$ as well: these are the ancillary qubits. In other words, on every iteration, we double the number of columns. In particular, on an iteration for which \eqref{changepar} is false, the reordering takes each column (whose length has been doubled by the addition of the new computational qubit) and splits it into two columns, each of which has the same length as the original column. On an iteration for which \eqref{changepar} is true, the reordering step requires two additional qubits: each column still splits into two columns, but each of these is 4 times as long as the original column, reflecting the fact that the number of slots and the sizes of the slots have both doubled.

Summing \eqref{npar}, \eqref{nstat},  and \eqref{nseq} gives us the total number of qubits used: after some simplification we obtain
\begin{equation}
	\text{total number of qubits (exact space requirement)}=n+2\lfloor\log_2(n)\rfloor-1,
\end{equation}
i.e., the computation requires only $2\lfloor\log_2(n)\rfloor-1$ ancillary qubits.

\noindent\textbf{Example:}

Consider the iteration $n=4\rightarrow5$. \eqref{changepar} is false, so we expect to only add the new computational qubit. Our input state has the form \eqref{inputvector}. We show the three pieces of the iteration step for the first column only, since the action will be the same on the other columns, just with fewer active slots:
\begin{equation}
	\label{itstep}
	\hspace{0in}
	\vcenter{\includegraphics[height=1.5in]{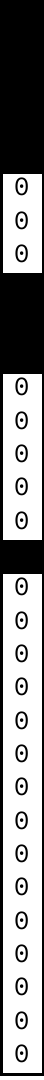}}
	\hspace{-5.3in}\stackrel{\text{tensor in new qubit}}{\longrightarrow}\hspace{-0.17in}
	\vcenter{\includegraphics[height=2in]{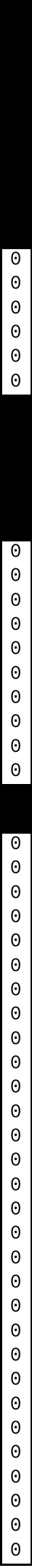}}
	\hspace{-5.3in}\stackrel{\text{apply super-CG transform}}{\longrightarrow}\hspace{-0.17in}
	\vcenter{\includegraphics[height=2in]{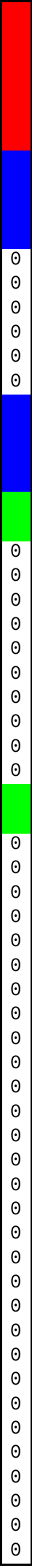}}
	\hspace{-5.3in}\stackrel{\text{reorder by new irreps}}{\longrightarrow}\hspace{-0.1in}
	\vcenter{\includegraphics[height=2in]{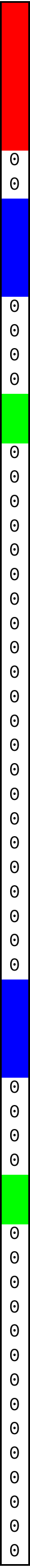}}
	\hspace{-5.2in}=\hspace{-0.1in}
	\vcenter{\includegraphics[height=1.5in]{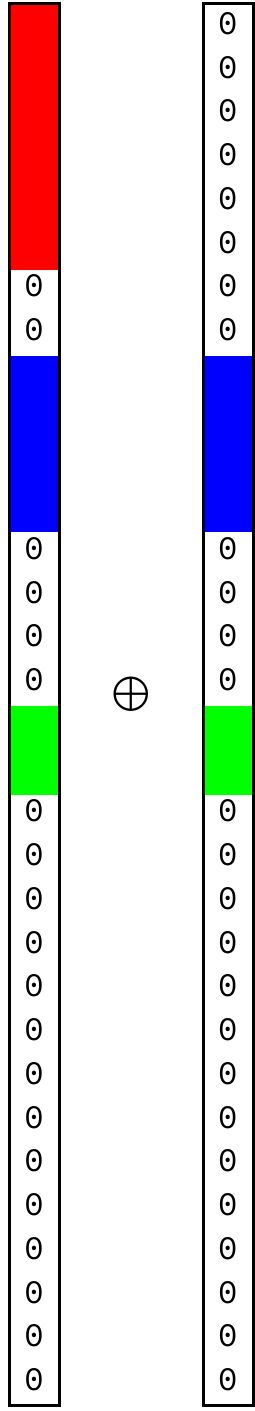}}
\end{equation}
In the first step, we add the new computational qubit, doubling the dimensions of the slots. In the second step, we apply the appropriate CG transform to each slot, which splits the tensor product of each original irrep and the new qubit into two new irreps. We have labeled the new irreps by color:
\begin{align}
	\label{outputcolors}
	&\mathcal{Q}_{(5,0)}^2:\quad\text{red in \eqref{itstep}}\nonumber\\
	&\mathcal{Q}_{(4,1)}^2:\quad\text{blue in \eqref{itstep}}\nonumber\\
	&\mathcal{Q}_{(3,2)}^2:\quad\text{green in \eqref{itstep}}
\end{align}
In terms of spins, our original irreps were spin-$2$, spin-$1$, and spin-$0$. We can see in \eqref{itstep} that the CG step splits the spin-2 into a spin-$(5/2)$ ($\mathcal{Q}_{(5,0)}^2$) and a spin-$(3/2)$ ($\mathcal{Q}_{(4,1)}^2$), splits the spin-1 into a spin-$(3/2)$ and a spin-$(1/2)$ ($\mathcal{Q}_{(3,2)}^2$), and simply lifts the spin-0 to a spin-$(1/2)$. We then reorder the basis according to these new spin-subspaces. Notice that since there are still only three inequivalent irreps, and the highest dimension is now 6 instead of 5, we still need only 2 \textbf{par} qubits (4 slots) and 3 \textbf{stat} qubits (8 states within each slot); thus, as we expected, the columns remain the same size.

\noindent\rule{3in}{0.4pt}

By repeatedly applying the iteration step described above, we obtain the Schur transform via a product of super-CG and reordering transforms. Each super-CG and reordering transform pair is tensored into the identity matrix of appropriate dimension to copy it over the states of \textbf{seq}. We complete our algorithm by decomposing each super-CG transform and reordering transform directly into a product of Clifford+T gates, using known methods for general unitary decomposition \cite{kliuchnikov13,kliuchnikov13b,selinger14}. We will discuss this decomposition in more detail in the next section, in which we determine the resulting sequence length.

\section{Analysis for qubits}

\label{analysis}
The purpose of the ordering of the basis discussed in \cref{ordering} and \cref{iterationstep} is to allow the action of the iteration step to be copied over the columns, that is, over the states of \textbf{seq}. We now show how this allows our algorithm for qubits to be efficient.

Our algorithm decomposes the Schur transform on $n$ qubits into $O(n^3)$ two-level unitary operators (unitaries that only act nontrivially on two dimensions). Each of these can be approximated by a sequence of operators from the Clifford+T set \cite{gottesman98,bravyi05}, which can be implemented fault-tolerantly. We break down the number of two-level unitaries required to construct our algorithm as follows:
\begin{itemize}
	\item The Schur transform on $n$ qubits requires $n-1$ iteration steps, since the Schur basis is identical to the computational basis for 1 qubit.
	
	\item For the iteration step that lifts $k\rightarrow k+1$ qubits, the super-CG transform is block diagonal, with each block a CG transform corresponding to the addition of $1$ qubit to each inequivalent irrep for $k$ qubits. In particular, for $\mu=(\mu_1,\mu_2)\vdash k$, the block associated to the addition of $1$ qubit to $\mathcal{Q}_\mu^2$ has side length equal to twice the dimension of $\mathcal{Q}_\mu^2$. Thus by \eqref{dimq}, the side length of the block corresponding to $\mathcal{Q}_\mu^2$ is
		\begin{equation}
			2(\mu_1-\mu_2+1)
		\end{equation}
	
	\item We can reduce a nonsingular square matrix to upper-triangular using one two-level rotation per nonzero entry below the main diagonal \cite{bernstein05}. If we reduce a unitary matrix to upper-triangular, we must have reduced it to a diagonal matrix whose diagonal entries have norm $1$, since this is the form of any upper-triangular unitary matrix. By then applying a one-level phase shift (which we are free to think of as a two-level rotation) to each main diagonal entry, we can map that diagonal matrix to the identity matrix. We refer to the total number of two-level rotations into which the goal matrix is decomposed as the \emph{two-level gate sequence length}.
	
	\item The number of nonzero entries on or below the main diagonal thus gives us our two-level gate sequence length for a single CG transform. This sum is bounded by the total number of nonzero elements in the CG transform (and is of the same order). Each row in the CG transform is an eigenvector of the total spin projection: suppose a particular row $\langle\phi|$ has total spin projection $m$. Then for any computational basis vector $|m_1,m_2\rangle$ appearing in the linear combination that forms $|\phi\rangle$, $m_1+m_2=m$. But since our CG transforms always just add a single qubit to some irrep $\mathcal{Q}_\mu^2$, $m_2$ is restricted to the values $\pm\frac{1}{2}$. Therefore, at most two computational basis vectors appear in the linear combination that forms $|\phi\rangle$, i.e., there are at most two nonzero entries per row in the CG transform. Therefore, the two-level gate sequence length for the CG transform is bounded by $2\ell$, where $\ell$ is the side length of the CG transform.
	
	\item In our case, $\ell=2(\mu_1-\mu_2+1)$, so the sequence length for the CG transform on $\mathcal{Q}_\mu^2$ is bounded by
		\begin{equation}
			4(\mu_1-\mu_2+1)=4(k-2\mu_2+1)
		\end{equation}
		for $\mu_1+\mu_2=k$.
		
	\item The super-CG transform is a direct sum of CG transforms on $\mathcal{Q}_\mu^2$, for every $\mu\vdash k$ with degree $2$. The possible $\mu$ are
		\begin{equation}
			\mu\in\left\{(k-\mu_2,\mu_2)\ |\ \mu_2\in\left\{0,1,2,...,\left\lfloor\frac{k}{2}\right\rfloor\right\}\right\}
		\end{equation}
		Therefore, the sequence length for the super-CG transform is
		\begin{equation}
			\label{itstepanalysis}
			4\sum_{\mu_2=0}^{\left\lfloor k/2\right\rfloor}(k-2\mu_2+1)
		\end{equation}
		
	\item Our Schur transform decomposes into super-CG transforms and reordering transforms for $k=1,2,...,n-1$, as discussed above; so far we have only discussed the sequence lengths for the super-CG transforms. Each reordering transform is a permutation matrix that permutes exactly the same entries that its corresponding super-CG transform acts on. Thus, it has at most 1 off-diagonal element for each of these entries. Since the super-CG transform had at most 2, the effect of including the reordering transform in our analysis is simply to increase the constant factor in \eqref{itstepanalysis} from 4 to 6. Thus, the two-level gate sequence length for the full Schur transform is bounded by
		\begin{equation}
			\label{runtime}
			6\sum_{k=1}^{n-1}\sum_{\mu_2=0}^{\left\lfloor k/2\right\rfloor}(k-2\mu_2+1)\le\frac{1}{3}n^3+\frac{9}{4}n^2+\frac{13}{4}n-6<2n^3
		\end{equation}
		for $n\ge2$; this is $O(n^3)$, as desired.
\end{itemize}

We can test our prediction for the two-level gate sequence length directly. In Fig.\ 4, the black line is $n^3$, and the solid points are the actual two-level gate sequence lengths generated by our algorithm for the Schur transform. As it turns out, the actual two-level gate sequence length appears to be bounded by $n^3$; the additional factor of $2$ in \eqref{runtime} arises as a result of allowances we made for the sake of simplicity in our analysis.

It also behooves us at this point to note that one does not obtain efficient sequence lengths by direct decomposition of the Schur transform into two-level gates using a method such as that described in \cite{kliuchnikov13}, i.e., by bypassing our algorithm entirely. We expect the sequence lengths thus generated to be exponential in scaling, and this is indeed what we obtain: the open circles in Fig.\ 4 are the two-level gate sequence lengths generated by decomposing the Schur transform directly.
\begin{figure}[h!]
	\caption{Two-level gate sequence lengths for $n=2\text{ to }20$}
	\centering
	\includegraphics[width=4in]{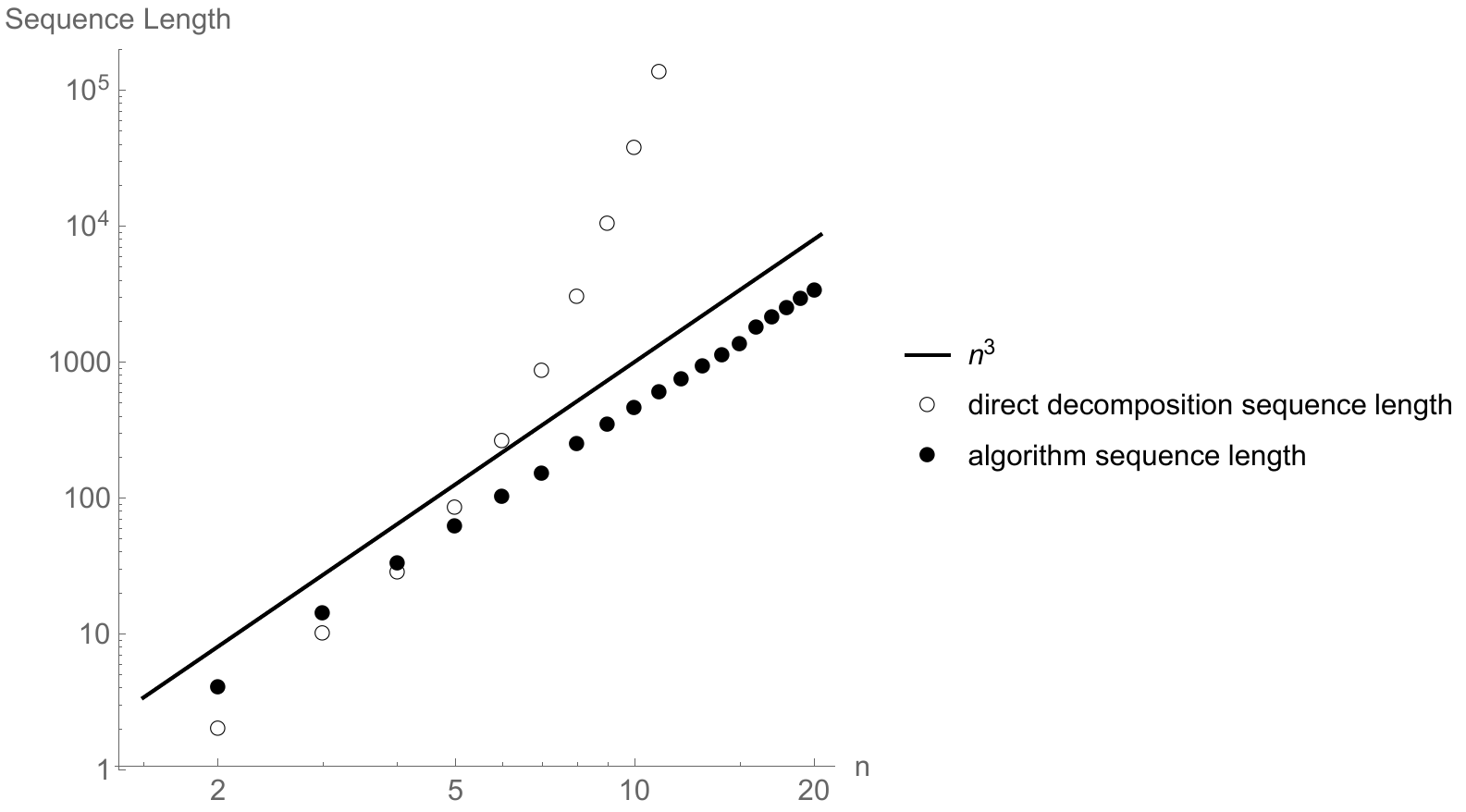}
\end{figure}

Now that we have our sequence of $O(n^3)$ two-level rotations, we can approximate each two-level rotation by Clifford+T operators using known methods \cite{kliuchnikov13,kliuchnikov13b,selinger14}: a two-level rotation on $n$ qubits can be decomposed into $O(n\log(1/\delta))$ Clifford+T operators, with error $\delta$. If we want to achieve an overall error bounded by $\eps$, we can calculate the error bound required for the individual two-level rotations as follows:

\begin{lemma}
\label{errorlemma}
Let $A_1,A_2,...,A_m$ be a sequence of unitary operators. If $B_1,B_2,...,B_m$ is a sequence of unitary operators such that $B_j$ approximates $A_j$ with error $\delta$ in the trace norm for all $j$, then the product $\Pi_{j=1}^mB_j$ approximates the product $\Pi_{j=1}^mA_j$ with error $\eps\le m\delta$ \cite{nielsen01}.
\end{lemma}

Therefore, if we want our overall error to be bounded by $\eps$, the errors for our decomposition matrices (the two-level rotations) must be bounded by $\delta=\eps/m$. We decompose into $m=O(n^3)$ two-level rotations, so our decomposition errors for the individual rotations must be bounded by $\frac{\eps}{a n^3}$ for some scalar $a$. Thus the lengths of the approximating sequences will be bounded by
\begin{equation}
	O\left(n\log\left(\frac{a n^3}{\eps}\right)\right)=O\left(3n\log\left(\frac{n}{\eps}\right)\right)=O\left(n\log\left(\frac{n}{\eps}\right)\right)
\end{equation}
Multiplying by the number of two-level gates in our decomposition gives us the overall sequence length for decomposition of the Schur transform on $n$ qubits into Clifford+T operators:
\begin{equation}
	\label{qubitruntime}
	O\left(n^4\log\left(\frac{n}{\eps}\right)\right)
\end{equation}
We show in \cref{discussionc} that this length agrees with that of the circuit schematically described in \cite{bacon06}.

We have implemented this method of calculating an efficient quantum algorithm for the qubit Schur transform both as a Python script and in Mathematica: the code is available online at \cite{schurgithub}. We used our implementation to generate the sequence lengths in Fig.\ 4, and also tested it for correctness by multiplying out the resulting Schur transforms and checking that
\begin{enumerate}
	\item they match previous calculations,
	\item they diagonalize the global spin operators, and
	\item they reduce the representations of $\mathcal{U}_2$ and $\mathcal{S}_n$ described in \cref{mathbackground} to direct sums of irreps.
\end{enumerate}

\section{Implementation and analysis for qudits}
\label{qudit}
We now generalize from qubits to qudits of dimension $d$. Up to and including the decomposition into two-level gates, the only differences will be in the dimensions, numbers, and multiplicities of the irreps, as well as in the CG transforms used. The similarity ends with the decomposition of the two-level rotations into primitives from a universal gate set. The Clifford+T gate set is specific to qubits, so in order to complete this piece of the construction, we need a different universal gate set that applies to general qudits. Ideally, this gate set would be fault-tolerant, and the decomposition of a two-level rotation into primitives from the gate set would be $O(n\log(1/\eps))$. To our knowledge, no constructive algorithm that satisfies these conditions has been found. However, Gottesman showed that fault-tolerant computation with qudits is possible, and provided a generalization of the Clifford+T set that is universal and fault-tolerant \cite{gottesman99}. Thus, we can apply the Solovay-Kitaev algorithm \cite{kitaev02,dawson05} to this set to obtain a fault-tolerant decomposition of any two-level rotation into $O(n\log^p(1/\eps))$ primitives, for $p\approx3.97$. This will be sufficient to show that our construction gives a efficient decomposition of the Schur transform into primitives; the only further improvements would be in the dependence on $\eps$, for which $p$ could in principle be decreased to $1$ (note: this has been done \cite{wang15}, but not fault-tolerantly).

We can break down the number of two-level rotations required to construct our algorithm for the Schur transform on $n$ qudits (dimension $d$) as follows:

\begin{itemize}
	\item The Schur transform on $n$ qudits requires $n-1$ super-CG transforms.
	
	\item The super-CG transform that adds $1$ qudit to $k$ qudits is block diagonal, with each block corresponding to the addition of $1$ qudit to an irrep of $k$ qudits. In particular, for $\mu=(\mu_1,\mu_2,...,\mu_d)\vdash k$, the block associated to the addition of $1$ qudit to $\mathcal{Q}_\mu^d$ has side length equal to $d$ times the dimension of $\mathcal{Q}_\mu^d$. Note that $\mu$ cannot have degree greater than $d$, so we can write $\mu=(\mu_1,\mu_2,...,\mu_d)$ as long as we remember that $\mu_2$ through $\mu_d$ may be $0$.
	
	\item The dimension of $\mathcal{Q}_\mu^d$ is given by the number of semistandard $\mu$-tableaux $t$ with entries in $\{1,2,...,d\}$ \eqref{dimq}. For $\mu\vdash k$, $\mu$ has the maximal number of semistandard tableaux if $\mu=(k)$, as was the case for qubits. The number of semistandard tableaux with shape $(k)$ and entries in $\{1,2,...,d\}$ is bounded by~\cite{christandl06}
    \begin{equation}
        (k+1)^{d(d-1)/2}.
    \end{equation}
		
	\item The number of distinct irreps is given by the number of distinct partitions of $k$ with order bounded by $d$. This number is bounded by $k^d$, since we choose the length of each of the $d$ rows from at most $k$ possibilities.
	
	\item Our super-CG transform on $k$ qudits will have a block corresponding to each distinct irrep of $k$ qudits. The side length of each of these blocks is $d$ times the dimension of the corresponding irrep, so the number of two-level rotations required to implement the block is bounded by the square of this side length. Since the dimension of any irrep is bounded by $(k+1)^{d(d-1)/2}$ for $k$ qudits, the side length of the corresponding block is bounded by $d(k+1)^{d(d-1)/2}$. Thus, since the number of distinct irreps is bounded by $k^d$, and hence also by $(k+1)^d$, the number of nonzero entries in the super-CG is bounded by
		\begin{equation}
			d^2(k+1)^{d(d-1)}\cdot(k+1)^d=d^2(k+1)^{d^2}
		\end{equation}
		This is a bound on the number of two-level rotations to decompose the super-CG transform.
	
	\item To build the Schur transform, we take the product of super-CG transforms on $k$ qudits for $k$ from $1$ to $n-1$, so the total number of two-level rotations to decompose the Schur transform on $n$ qudits is bounded (for fixed $d$) by
		\begin{equation}
			d^2\ \sum_{k=1}^{n-1}(k+1)^{d^2}\le d^2\int_{k=1}^n(k+1)^{d^2}dk<(n+1)^{d^2+1}<O(n^{d^2+1})
		\end{equation}
		As discussed above, each two-level rotation can be decomposed to accuracy $\delta$ into $O(n\log^p(1/\delta))$ fault-tolerant primitives (for $p\approx3.97$), so the Schur transform can be decomposed into a product of fault-tolerant primitives whose length is
		\begin{equation}
			O\left(n^{d^2+2}\log^p\left(\frac{n^{d^2+1}}{\eps}\right)\right)
		\end{equation}
		for overall error bounded by $\eps$. Since this is polynomial in $n$, our goal is achieved.
	
\end{itemize}

A tighter analysis than we have given is certainly possible, but ours is sufficient for proof of principle.

\section{Discussion}
\label{discussionc}
It is appropriate for us to conclude by comparing the algorithms presented here with those of BCH \cite{bacon07,bacon06} and summarizing what we have contributed to the study of the quantum Schur transform. The following is an itemization of the similarities and differences between these algorithms:
\begin{enumerate}

	\item BCH's algorithm and ours each implement the Schur transform on $n$ qubits or qudits.
	
	\item BCH's algorithm and ours share the same recursive structure at the outermost level, but ours uses a different structure for qudits at lower levels (expanding upon arguments given in Section V in \cite{bacon07}).  In this sense we have extended the BCH algorithm.
		
	\item Our algorithm has low prefactor overhead in terms of the decomposition into two-level gates: for qubits this sequence length is bounded by $2n^3$, and for qudits it is bounded by $d(n+1)^{d^2+1}$.
		
	\item For the case of qubits, the asymptotic sequence length of the BCH algorithm in terms of Clifford+T gates is $O(n\,\text{polylog}(n,1/\epsilon))$, compared to $O\left((n^4\,\log\left(\frac{n}{\epsilon}\right)\right)$ for our construction. 
	
	\item Our algorithm employs exactly $2\lfloor\log_2(n)\rfloor-1$ ancillary qubits.  A direct implementation of the BCH algorithm would use $n$ ancillary qubits (although these can be compressed afterwards \cite{bacon07}).
	
	\item For the case of qudits, we provide an explicit asymptotic sequence length that is polynomial in $n$; BCH specify that their algorithm is polynomial in $n$ and $d$ but do not state the exponents.  However, in many proposed quantum computation architectures, $d$ is a constant with $n$ the only variable.
		
	\item The main advantage of BCH's qudit algorithm is that it is polynomial in $d$, while our qudit algorithm is exponential in $d$.  This comes at the cost of constructing the required Clebsch-Gordan coefficients by using a reduced Wigner operator (via the Wigner-Eckart Theorem).  	
		
	\item The main advantage of our qudit algorithm is its relative simplicity. Our algorithm includes only the most basic operations required to implement the Schur transform. BCH employ additional mathematical machinery to reduce their sequence length's asymptotic dependence on $d$ to polynomial. Part of our contribution is to show that if we relax that requirement, with the understanding that in many quantum computers the asymptotic dependence on $d$ will be irrelevant, we can obtain an algorithm that is efficient in $n$ using only the most basic mathematical components.
	
	\item These most basic mathematical components are the CG transforms. At each step in the recursive structure of our algorithm (or BCH's), we must implement a CG transform to act on each irrep output from the previous step. The CG transforms acting on equivalent irreps are identical, so the minimum number of operations for the step is to implement each of these once; this is exactly what we accomplish.
	
	\item To complete the Schur transform all we must do is route the correct outputs from one step into the correct inputs from the next. The calculation of CG transforms is a solved problem: CG coefficients could be looked up from a precalculated table if we desired (e.g., using \cite{alex11}), and doing so would not change our sequence length. Thus, in a real sense we have reduced the Schur transform to a succession of permutations of the basis that implement the ``routing."  In this sense also we have simplified the BCH algorithm.
		
	\item The remainder of our contribution is to provide an explicit implementation of the qubit version of the algorithm, described in \cref{qubit} and also available online \cite{schurgithub}, which allows us to verify our results (see \cref{analysis}). Given the various applications for the Schur transform, it is our hope that this implementation and our presentation will be useful to others approaching quantum computation and the Schur transform from a computer science or physics background.
	
\end{enumerate}

\nonumsection{Acknowledgements}
\noindent
We thank William Wootters, Adam Wills, and Dmitry Grinko for their helpful discussions. We are also extremely grateful to Dmitry and Adam for pointing out errors in prior versions.

\bibliographystyle{apsrev}
\bibliography{library.bib}

\end{document}